\documentclass{iopart}
\usepackage{iopams}

\expandafter\let\csname equation*\endcsname\relax
\expandafter\let\csname endequation*\endcsname\relax
\usepackage{amsmath}
\usepackage{xcolor}
\usepackage{tikz}
\usepackage{tikz-cd} 
\usetikzlibrary{decorations.markings,arrows}
\usepackage{multirow}
\usepackage{booktabs}

\usepackage[color=green!60]{todonotes}
\usepackage[normalem]{ulem}

\usepackage{color}
\usepackage{graphicx}
\usepackage{amsthm}
\usepackage{bbm}
\usepackage{verbatim}
\usepackage{physics}
\DeclareMathAlphabet{\mathpzc}{OT1}{pzc}{m}{it}

\begin{document} 
\title[Moments of the inverse participation ratio]
{Moments of the inverse participation ratio for the Laplacian on finite regular graphs}
\author{Timothy B. P. Clark} 
\address{Department of Mathematics and Statistics,
       Loyola University Maryland, Baltimore, MD 21211, USA}
\ead{tbclark@loyola.edu}

\author{Adrian {Del Maestro}}
\address{Department of Physics, University of Vermont, Burlington, VT 05405,
USA}
\ead{Adrian.DelMaestro@uvm.edu}

\date{\today}

\newtheorem{intheorem}{Theorem} 
\newtheorem{inlemma}[intheorem]{Lemma} 
\newtheorem{inproposition}[intheorem]{Proposition} 
\newtheorem{incorollary}[intheorem]{Corollary}
\newtheorem{indefinition}[intheorem]{Definition} 
\newtheorem{inremark}[intheorem]{Remark} 
\newtheorem{innotation}[intheorem]{Notation}

\newtheorem{theorem}[equation]{Theorem} 
\newtheorem{lemma}[equation]{Lemma} 
\newtheorem{proposition}[equation]{Proposition} 
\newtheorem{corollary}[equation]{Corollary} 
\newtheorem{observation}[equation]{Observation} 
\newtheorem{conjecture}[equation]{Conjecture} 
\newtheorem{fact}[equation]{Fact} 
\newtheorem{example}[equation]{Example} 
\newtheorem{definition}[equation]{Definition} 
\newtheorem{remark}[equation]{Remark} 
\newtheorem{remarks}[equation]{Remarks} 
\newtheorem{notation}[equation]{Notation} 

\renewcommand{\:}{\! :\ } 
\newcommand{\p}{\mathfrak p} 
\newcommand{\m}{\mathfrak m}
\newcommand{\g}{{\bf g}} 
\newcommand{\lra}{\longrightarrow} 
\newcommand{\ra}{\rightarrow} 
\newcommand{\altref}[1]{{\upshape(\ref{#1})}} 
\newcommand{\bfa}{\mathbf a} 
\newcommand{\bfb}{\boldsymbol{\beta}} 
\newcommand{\bfg}{\boldsymbol{\gamma}} 
\newcommand{\bfd}{\boldsymbol{\delta}} 
\newcommand{\bfM}{\mathbf M} 
\newcommand{\bfN}{\mathbf N}
\newcommand{\bfI}{\mathbf I} 
\newcommand{\bfC}{\mathbf C} 
\newcommand{\bfB}{\mathbf B} 
\newcommand{\mcP}{\mathcal P}
\newcommand{\mcX}{\mathcal X}
\newcommand{\ipr}{\ensuremath{{IPR}}}
\newcommand{\D}{\textbf{\textup{D}}}
\newcommand{\bsfC}{\bold{ C}} 
\newcommand{\bsfT}{\bold{\mathsf T}}
\newcommand{\mc}{\mathcal} 
\newcommand{\smsm}{\smallsetminus} 
\newcommand{\ol}{\overline} 
\renewcommand{\vec}[1]{\boldsymbol{#1}}
\newcommand{\twedge}
           {\smash{\overset{\mbox{}_{\circ}}
                           {\wedge}}\thinspace} 
\newcommand{\mbb}[1]{\mathbb{#1}}
\newcommand{\pring}{\Bbbk[x_1,\ldots,x_d]}
\newcommand{\irr}{(x_1,\ldots,x_d)}
\newcommand{\Z}{\textup{Z}}
\newcommand{\B}{\textup{B}}
\newcommand{\La}{\mathcal{L}}
\newcommand{\redHo}{\widetilde{\textup{H}}}
\newcommand{\redH}[2]{\widetilde{\textup{H}}_{#1}(#2)}
\newcommand{\pr}{\textup{proj}}
\newcommand{\precdot}{\prec\!\!\!\cdot\;}
\newcommand{\succdot}{\;\cdot\!\!\!\succ}
\newcommand{\vset}[2]{\textbf{V}_{{#1},{#2}}}
\newcommand{\sx}[2]{\textbf{B}_{{#1},{#2}}}
\newcommand{\cplx}[2]{\Delta_{{#1},{#2}}}
\newcommand{\vx}[1]{(\varnothing,#1)}
\newcommand{\mbf}[1]{\mathbf{#1}}
\newcommand{\ds}{\displaystyle}
\newcommand{\Sy}{\Sigma}
\newcommand{\Ho}{\widetilde{H}}
\newcommand{\sgn}{\textup{sgn}}
\newcommand{\CC}{\widetilde{\mathcal{C}}}
\newcommand{\ld}{\lessdot}
\newcommand{\mdeg}{\textup{mdeg}}
\newcommand{\lcm}{\textup{lcm}}
\newcommand{\id}{\textup{id}}
\newcommand{\BigFact}{\mbox{\Large{!}}}
\newcommand{\Gamfunc}[2]{\Gamma\qty(\frac{#1}{#2})}

\newlength{\wdtha}
\newlength{\wdthb}
\newlength{\wdthc}
\newlength{\wdthd}
\newcommand{\elabel}[1]
           {\label{#1}  
            \setlength{\wdtha}{.4\marginparwidth}
            \settowidth{\wdthb}{\tt\small{#1}} 
            \addtolength{\wdthb}{\wdtha}
            \smash{
            \raisebox{.8\baselineskip}
            {\color{red}
             \hspace*{-\wdthb}\tt\small{#1}\hspace{\wdtha}}}}

\newcommand{\mlabel}[1] 
           {\label{#1} 
            \setlength{\wdtha}{\textwidth}
            \setlength{\wdthb}{\wdtha} 
            \addtolength{\wdthb}{\marginparsep} 
            \addtolength{\wdthb}{\marginparwidth}
            \setlength{\wdthc}{\marginparwidth}
            \setlength{\wdthd}{\marginparsep}
            \addtolength{\wdtha}{2\wdthc}
            \addtolength{\wdtha}{2\marginparsep} 
            \setlength{\marginparwidth}{\wdtha}
            \setlength{\marginparsep}{-\wdthb} 
            \setlength{\wdtha}{\wdthc} 
            \addtolength{\wdtha}{1.1ex}
            \marginpar{\vspace*{-0.3\baselineskip}
                       \tt\small{#1}\\[-0.4\baselineskip]\rule{\wdtha}{.5pt} }
            \setlength{\marginparwidth}{\wdthc} 
            \setlength{\marginparsep}{\wdthd}  }  
            
\renewcommand{\mlabel}{\label} 
\renewcommand{\elabel}{\label} 

\newcommand{\mysection}[1]
{\section{#1}\setcounter{equation}{0}
             \numberwithin{equation}{section}}

\newcommand{\mysubsection}[1]
{\subsection{#1}\setcounter{equation}{0}
                \numberwithin{equation}{subsection}}

\newcommand{\mysubsubsection}[1]
{\subsubsection{#1}\setcounter{equation}{0}
                \numberwithin{equation}{subsubsection}}

\begin{abstract}
We investigate the first and second moments of the inverse participation ratio (IPR) for all eigenvectors of the Laplacian on finite random regular graphs with $n$ vertices and degree $z$.  By exactly diagonalizing a large set of $z$-regular graphs, we find that as $n$ becomes large, the mean of the inverse participation ratio on each graph, when averaged over a large ensemble of graphs, approaches the numerical value $3$.  This universal number is understood as the large-$n$ limit of the average of the quartic polynomial corresponding to the IPR over an appropriate $(n-2)$-dimensional hypersphere of $\mathbb{R}^n$. For a large, but not exhaustive ensemble of graphs, the mean variance of the inverse participation ratio for all graph Laplacian eigenvectors deviates from its continuous hypersphere average due to large graph-to-graph fluctuations that arise from the existence of highly localized modes. 
\end{abstract}

\maketitle 

\section{Introduction} 
Much of condensed matter physics involves the study of either localized or
itinerant degrees of freedom that exist on the sites of a Euclidean lattice,
defined by a notion of physical distance between a site and some number of
proximate or ``neighboring'' ones.  The relationship between a site and its
neighbors defines both the dimension of space, $d$ and a finite set of lattice
vectors $a_j \in \mathbb{R}^d$, $j=1,\ldots,d$ that can be used in
combination with a set of integers $\{\vec{m}\}$ with $\vec{m}_i \in
\mathbb{Z}^d$ to index $n$ lattice sites via $\vec{R}_i = \sum_{j=1}^{d} m_{ij}
a_j$. Examples include the fourteen Bravais lattices in three spatial
dimensions.  However, it is often instructive to consider the same physical
degrees of freedom on a non-Euclidean lattice, or graph, where no distance
metric exists. A finite lattice of $n$ sites is replaced by a graph $G$,
consisting of a set of vertices $V=\{v_i\}$, each connected by $z_i$ undirected
edges to its neighboring vertices.  The quantity $z_i$ is known as the
\emph{degree} of the vertex $v_i$, and is equivalent to twice the spatial
dimension for a hypercubic lattice. 

Studying a given physical model on a graph offers many technical benefits
including the ability to: (i) study arbitrarily long range interactions where
exact mean-field solutions may be available, (ii) smoothly tune and control the
local dimensionality and (iii) easily encode the randomness and disorder that
often exists in real systems. Celebrated examples from statistical physics
include the solution of the Ising model and its generalizations; graph coloring
and random percolation problems (for a review see Ref.~\cite{Essam:1971ig}).
Anderson's model of non-interacting electrons hopping on a disordered lattice
\cite{Anderson:1958uc} was first solved on the Cayley tree (Bethe lattice)
\cite{AbouChacra:1973sc}, providing deep physical insights into the nature of
localization in quantum mechanical systems. More recently, the ability to study
graphs in the limit $z\to\infty$ has lead to the development of Dynamical Mean
Field Theory \cite{Georges:1996dm}, allowing for systematic investigations of
candidate microscopic models of the high temperature superconductors
\cite{Lichtenstein:2000ad}.

The discrete Laplacian matrix $L$ plays a crucial role in defining
any physical model on a graph, as it quantifies the energy cost of
rapidly varying some local degree of freedom among a set of neighboring vertices.
For example, it encodes the classical dynamics in random vibrational networks
\cite{Hastings:2003wr} as well the onset of ferromagnetism in the classical
\cite{Burioni:2000ds} and quantum $\mathrm{O}(n)$ model \cite{Laumann:2009gr,
Murray:2012cd} on graphs.  It appears in models of non-interacting bosons 
hopping between graph vertices, where the existence of a Bose-Einstein
condensation transition on complex networks can be rigorously proven
\cite{Burioni:2001in}.

\begin{figure}[ht]
\begin{center}
\includegraphics[width=0.3645\columnwidth]{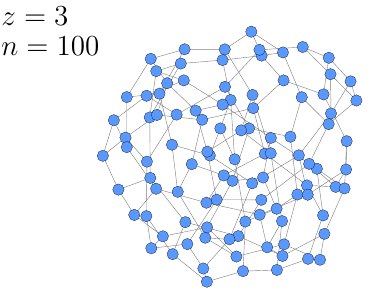}
\includegraphics[width=0.340\columnwidth]{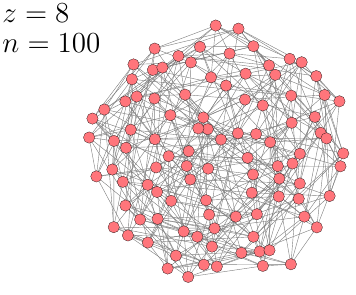}
\end{center}
\caption{Regular graphs with $n = 100$ vertices and degrees $z=3$ and
    $z=8$ constructed using \texttt{NetworkX} \cite{Hagberg:2008en}  via the
Steger-Wormald algorithm \cite{Steger:1999vh,Kim:2003yl}.} 
\label{fig:rrgraph}
\end{figure}

In this paper, we are interested in the properties of the Laplacian defined on
finite sized \emph{regular} graphs, defined by the constraint that each of the
$n$ vertices is connected to exactly $z$ neighbors.  Examples with $n=100$ and
$z=3,8$ are shown in Fig.~\ref{fig:rrgraph}.  
%
%
Such graphs possess many important mathematical properties \cite{Hoory:2006vo}
while still retaining similarities to the physically realizable Bravais
lattices discussed above. 

Much is known about the spectral properties of random regular graphs, both in
the thermodynamic limit $n\to\infty$
\cite{Kesten:1959rw,McKay:1981ex,Tran:2013sr} and more
recently, at finite (but large) $n$ \cite{Dumitriu:2012ge,
Metz:2014wj, Bauerschmidt:2016lk, Bauerschmidt:2017ua}. The analyses of spectral
statistics have yielded fruitful and universal connections \cite{Deift:2007uf}
between random regular graphs and the Gaussian Orthogonal Ensemble of random
matrix theory \cite{Jakobson:1999ky, Oren:2010je,Bauerschmidt:2017be} known to be relevant in
describing the fluctuations of energy levels in physical dynamical systems.%

Substantially less is known about the eigenvectors of $L$
\cite{Friedman:1993sg} with early work focusing on empirical
analyses of nodal domains \cite{Elon:2008it, Dekel:2010dz} or specific vectors
\cite{Kabashima:2012cu}, as as it is not possible to apply many of the
standard tools of analysis for Euclidean lattices, including the Fourier
transform.  Subsequently, a series of results
\cite{Dumitriu:2012ge,Brooks:2012ht,Geisinger:2013cd,Bauerschmidt:2017ua} have
shown that for suitably large $n$, the eigenvectors of random regular graphs are 
delocalized -- they have few non-zero entries. Very recently, the breakthrough
works of Bourgade, Huang, and Yau \cite{Bourgade:2017id} and Backhausz and
Szegedy \cite{Backhausz:2016ot} have proven exiting new results that the
eigenvector components are Gaussian independent and identically distributed for
large $n$. To our knowledge, the physical implications of these new
results for finite realizations of $z$ and $n$ amenable to direct numerical
analysis have yet to be explored.

To address this gap, we systematically study the eigenvectors of
the Laplacian matrix on a large class of finite size random regular graphs
through brute-force numerical diagonalization.  We investigate the statistics
of the inverse participation ratio, a scalar proxy for localization, and  
numerically observe that its mean across all eigenvectors approaches a finite
universal value equal to $3$, independent of graph degree.  We quantify the
second moment of the distribution and find a dependence both on degree $z$ and
the number of vertices $n$. 

The paper is organized as follows: we begin with a formal definition of the
discrete Laplacian on graphs in Section \ref{sec:the_graph_laplacian} and
describe our numerical results for the inverse participation ratio in Section
\ref{sec:inverse_participation_ratio}.  In Section \ref{sec:Theory}, we analyze
the inverse participation ratio as a polynomial function on an
$(n-2)$-dimensional subsphere of $n$-dimensional real space.  This perspective
allows us to calculate exact values for the first and second moments of the
inverse participation ratio over a continuous domain which contains the
(terminal points of the) eigenvectors of the Laplacian.  Section
\ref{sec:Experiments} compares the values of the theoretically derived moments
to those numerically computed from a large set of finite size random regular
graphs.  We analyze the deviations from the theoretically derived second moment
with a discussion of localized eigenvectors and highlight implications for
their use in computing physical observables on finite random regular graphs.

\section{The Graph Laplacian}
\label{sec:the_graph_laplacian}

The Laplacian matrix generalizes the continuous Laplace operator $\Delta
\equiv\vec{\nabla} \cdot \vec{\nabla}$ to encode variations of any continuous
function $\phi:V \to \mathbb{C}$ which can take a value $\phi_i$ on the 
vertex $v_i$. The physical importance of this matrix stems from the fact that
solutions of $\Delta \phi = 0$ correspond to the Dirichlet energy functional which is 
stationary in some spatial region.  The particular extension of $\Delta$ to a graph that
we employ arises from the discrete approximation to the second continuous
derivative of $\phi$ on a hypercubic lattice in $d$ spatial dimensions with
unit lattice spacing: 
\begin{equation}
    \Delta \phi (\vec{R}_i) \approx \sum_{j=1}^{d} 
    \left[\phi\left(\vec{R}_i + \vec{e}_j\right) + \phi\left(\vec{R}_i -
            \vec{e}_j\right) - 
    2 \phi\left(\vec{R}_i\right) \right]
\label{eq:hc_laplace}
\end{equation}
where $\vec{e}_j$ are the Cartesian unit vectors with elements $e_{kj} =
\delta_{kj}$ and $\delta_{kj}$ is the Kronecker $\delta$-function. On a
regular graph $G$ consisting of $n$ vertices $v_i$, each with degree $z$, the
local connectivity is encoded in an adjacency
matrix $A_{ij}$ where
\begin{equation}
    A_{ij} = \left\{
        \begin{array}{rcl}
            1 & ; & \mbox{if }v_i \mbox{ and } v_j \mbox{ share an edge}, \\
            0 & ; & \mbox{otherwise}.
        \end{array} \right.\, 
\label{eq:adjacency}
\end{equation}
Comparing with Eq.~(\ref{eq:hc_laplace}), we can write the elements of the
graph Laplacian matrix as the difference between the degree and adjacency
matrices of $G$:
\begin{equation}
    L_{ij} = z\delta_{ij} - A_{ij}
\label{eq:laplacian}
\end{equation}
and observe that $z$ corresponds to twice the dimension on a hypercubic
lattice.  As mentioned in the introduction, $L$ may appear in the Hamiltonian of
numerous physical systems defined on a graph in the form:
\begin{equation}
    H = \frac{1}{2} \sum_{i,j=1}^{n} \phi^{\ast}_i L_{ij} \phi^{\phantom\ast}_j.
\label{eq:hamiltonian}
\end{equation}
A spectral decomposition of $L$ provides a route to the determination of
the equations of motion governing a classical system, or the nature of the
wavefunctions and allowed energy eigenstates for a quantum mechanical one. 

\subsection{Exact diagonalization}
\label{ssub:Exact_diagonalization}

We now focus on the spectral decomposition of Laplacian
matrices drawn from an ensemble of random regular graphs with with $n$ vertices
and degree $z$. These matrices are generated using the $\mathrm{O}(n z^2)$
algorithm of Steger and Wormald \cite{Steger:1999vh}.  From the vertex neighbor
list of each graph, we construct the adjacency matrix $A$ and then exactly
diagonalize the resulting $n \times n$ sparse Laplacian matrix $L$ given in
Eq.~(\ref{eq:laplacian}).  In this paper, we present results for graphs with 
\begin{align*} z &\in \{3,4,5,10,15,20,25,30,35,40,45,50\} \\ n &\in \{200,
300, 400, 500, 1000, 2000, 3000, 4000, 5000, 10000\} \end{align*}
where $z$ and $n$ have been chosen with an eye towards
exploring their interdependence for large graphs. All averages are performed
over a set containing $N_G$ graph realizations, with $N_G = 5000$ for $n <
5000$ and $N_G=1000$ graphs for $n \ge 5000$.  The exact number of unique
graphs, $\mathcal{N}_G$ with a given $n$ and $z$ grows quickly with $n$ but is
unknown in general.  An asymptotic result for degrees satisfying $z\le
\sqrt{2\log n} - 1$ was proved by Bollob{\'a}s in Ref.~\cite{Bollobas:1980}.
Explicit counts for small $n$ and $z$ can be found at the Online Encyclopedia
of Integer Sequences \cite{OEIS}, \emph{e.g.} for $z=3$ and $n=16$,
$\mathcal{N}_G = 4060$.

We begin our analysis by describing the eigenvalue distribution of $L$. For $n
\gg 1$, the limiting form of the density of states $\rho(\varepsilon)$, the
probability of an eigenvalue falling between $\varepsilon$ and $\varepsilon +
d\varepsilon$, is given by the Kesten-McKay law
\cite{Kesten:1959rw,McKay:1981ex,Bauerschmidt:2016lk}:
\begin{equation}
    \rho(\varepsilon) = \frac{1}{n}\delta(\varepsilon) + \frac{z}{2\pi}\frac{\sqrt{4(z-1)-(\varepsilon-z)^2}}
    {z^2 - (\varepsilon-z)^2} \quad ; \quad |\varepsilon - z| \le 2\sqrt{z-1}.
\label{eq:rhoKM}
\end{equation}
For finite values of $n$, Metz~\emph{et al.} have recently computed
the $1/n$ corrections to this expression, originating from the contributions of
loops of all possible lengths.  

\begin{figure}[t]
\begin{center}
\includegraphics[width=1.00\columnwidth]{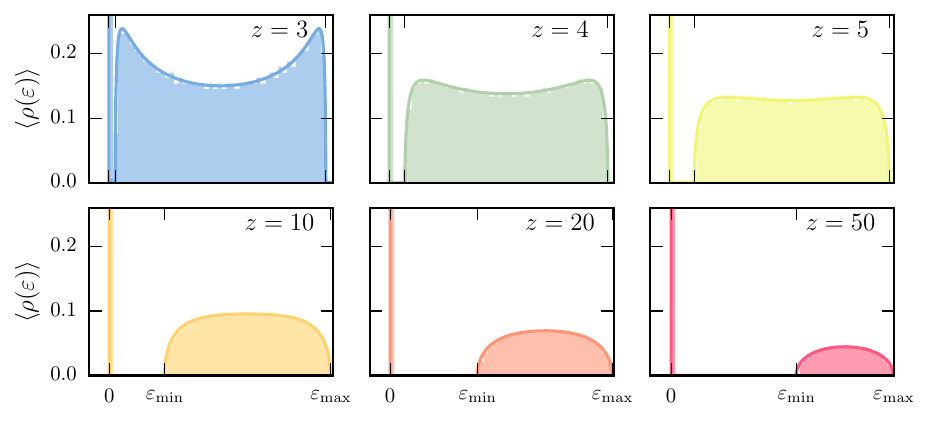}
\end{center}
\caption{The probability of graph Laplacian eigenvalues for $n =
1000$ vertices with degrees $z = 3,4,5,10,20,50$ computed by diagonalizing
numerically constructed graphs (shaded region) compared with the
large-$n$ limit Kesten-McKay law defined in Eq.~\eqref{eq:rhoKM} (solid line).
Eigenvalues in the continuum are bounded between $\varepsilon_{\mathrm{min}} =
z - 2\sqrt{z-1}$ and $\varepsilon_{\mathrm{max}} = z + 2\sqrt{z-1}$, while the
Perron-Frobenius mode with weight $1/n$ is shown as a spike at $\varepsilon = 0$.}
\label{fig:rrdistEvalsn} 
\end{figure}

For a graph with $n$ vertices, the eigenvalues of the Laplacian matrix 
$\{\varepsilon_i\}$ are determined by exact diagonalization, and
a comparison to Eq.~\eqref{eq:rhoKM} can be made by numerically constructing
the histogram: 
\begin{equation}
    \expval{\rho(\varepsilon)} = 
    \expval{\frac{1}{n}\sum_{i=1}^{n} \delta(\varepsilon-\varepsilon_i)}
\label{eq:evaluesdos}
\end{equation}
where an average over $N_G$ graph realizations is indicated
by the angle brackets: $\langle \cdots \rangle \equiv (1/N_G) \sum_{G} (
\cdots )$.  The results for $n=1000$,
$z=3,4,5,10,20,50$ and $N_G=5000$ are shown in Fig.~\ref{fig:rrdistEvalsn}.
%
%
We observe only small graph-to-graph variations and find excellent agreement
with the Kesten-McKay semi-circle law of Eq.~(\ref{eq:rhoKM}) using $50$
eigenvalue bins (solid line).  For $n > 1000$ there is no visible discrepancy
on this scale between the numerical results and the prediction for the large
$n$ limit.  For finite sized random regular graphs, the spectrum of the
Laplacian consists of a single eigenvalue at $\varepsilon=0$ separated by a
$z$-dependent gap \cite{Furedi:1981er} to a quasi-continuum of eigenvalues
bounded between $\varepsilon_{\mathrm{min}} = z - 2\sqrt{z-1}$ and
$\varepsilon_{\mathrm{max}} = z + 2\sqrt{z-1}$.

\section{The Inverse Participation Ratio}
\label{sec:inverse_participation_ratio}

\begin{figure}[ht]
\begin{center}
\includegraphics[width=1.00\columnwidth]{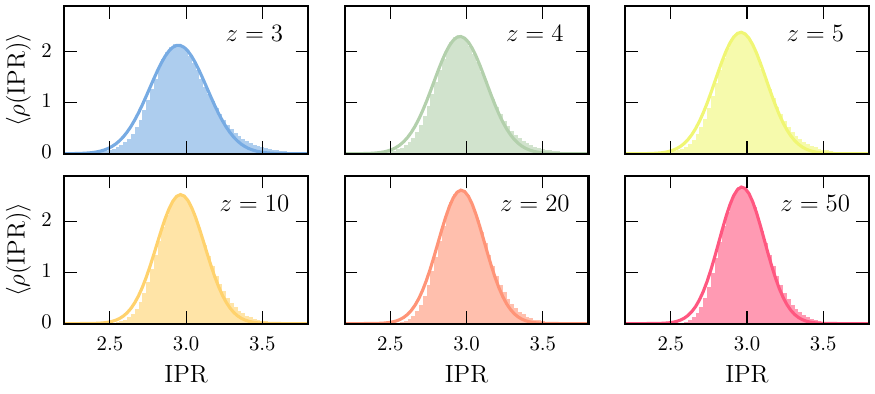}
\end{center}
\caption{Histograms of IPR values for the Laplacian matrix on random regular
graphs with $n = 1000$ vertices and degrees $z={3,4,5,10,20,50}$. Solid lines
represent fits to a Gaussian distribution.}
\label{fig:rrdistIPR}
\end{figure}

Less is understood about the set of eigenvectors $E=\{\vec{x} \,:\, L \vec{x} =
\varepsilon \vec{x} \}$ \cite{Elon:2008it, Dekel:2010dz, Tran:2013sr}
although for certain classes of regular graphs with $z \sim \mathrm{O}(n)$,
they are believed (with high probability \cite{Dumitriu:2012ge}) to be
delocalized -- meaning they have many non-zero components.  The eigenvalue
$\varepsilon=0$ with weight $1/n$ in Eq.~(\ref{eq:rhoKM}) and
Fig.~\ref{fig:rrdistIPR} corresponds to the special case of the
Perron-Frobenius mode: $\wp \equiv (1/\sqrt{n},\ldots,1/\sqrt{n})$ and via
orthogonality it follows that $\vec{x}\cdot\wp=\sum_i x_i=0$ for any
eigenvector $\vec{x}\ne \wp$.  We wish to develop an understanding of the
properties of the remaining eigenvectors $\mathcal{E}\equiv E\setminus
\{\wp\}$, and in particular, determine how the non-zero elements of an arbitrary  
$\vec{x} \in \mathcal{E}$ are distributed amongst its $n$ coordinates. 

To this end, we study the notion of \emph{localization} of an eigenvector 
using the inverse participation ratio. Historically, 
the participation ratio $p$ was introduced to aid in classifying the
properties of atomic vibrations in disordered lattices \cite{Bell:1970ic}. 
It describes the fraction of the total number of sites which \emph{participate} 
in a given normal mode vibration corresponding to the eigenvector
$\vec{x}=(x_1,\ldots,x_n)\in\mathbb{R}^n$ and takes the value 
\begin{equation}
    p(\vec{x}) = \frac{(\mu^1)^2}{\mu^0 \mu^2}
\label{eq:p}
\end{equation}
where $\mu^r = \sum_{i=1}^{n} |x_i|^{2r}$ can be thought of as the
$r^{\mathrm{th}}$ moment of the kinetic energy of the mode.  If a given normal
mode only involves the motion of a single atom, it is characterized as
\emph{localized} and has $p = 1/n$. A vibrational mode consisting of all
atoms participating equally is called \emph{extended} and has $p = 1$.
An equivalent measure was employed by Visscher \cite{Visscher:1972gg} to study
the degree of localization of electronic eigenstates in the Anderson model
\cite{Anderson:1958uc} with implications for the existence of a metal-insulator
(delocalization-localization) transition in the presence of disorder. 


When considering normalized eigenvectors $\vec{x} \cdot \vec{x} = || \vec{x} ||
= 1$, it is often more convenient to consider the associated inverse
participation ratio (IPR): 
\begin{equation}
    \frac{1}{p(\vec{x})} \equiv \ipr(\vec{x}) = n\sum_{i=1}^n x_i^4. 
\label{eq:ipr}
\end{equation}
For the Laplacian matrix in Eq.~(\ref{eq:laplacian})
with $\vec{x} \in E$ we have  
\begin{equation}
    1 \le \ipr(\vec{x}) \le \frac{n}{2}
    \label{eq:iprrange}
\end{equation}
with bounds corresponding to the extended Perron-Frobenius (lower bound) and localized
(upper bound) modes, respectively.  The finite size scaling of $\ipr(\vec{x})/n$ as $n \to
\infty$ provides information on the existence of a mobility edge, which defines
the portion of the spectrum with robust delocalized states. This scaling has been
extensively studied for a large class of random matrices \cite{Cizeau:1994bm,
Cavagna:1999hg, Metz:2010fd, Slanina:2012we}.

The inverse participation ratio thus provides a convenient single scalar value
measuring the degree to which a particular eigenvector is 
 localized ($p^{-1} \sim \mathrm{O}(n)$) or extended ($p^{-1} \sim
\mathrm{O}(1)$). To obtain information on the
reduced set of Laplacian eigenvectors $\mathcal{E}$ corresponding to non-zero
eigenvalues, we construct a histogram of values in analogy with
Eq.~(\ref{eq:evaluesdos}).  The non-linear form of the IPR necessitates that
the order of averaging is important, and we must compute
\begin{equation}
\rho(\mathrm{IPR})   =
\frac{1}{n-1}\sum_{\vec{x}\in\mathcal{E}}\delta(\mathrm{IPR}-\ipr(\vec{x}))
\end{equation}
for each graph realization separately before averaging over graphs.  The
resulting graph averaged distributions are shown in Fig.~\ref{fig:rrdistIPR}
for $n=1000$, $z=3,4,5,10,20,50$ and $N_G=5000$.
%
%
The solid lines show the results of a fit to a Gaussian distribution for each
$n$ and $z$ and there are clear deviations which skew to larger IPR values, most
notably for small $z$. For fixed $n$, increasing $z$ decreases the width of the
distribution, and slightly improves the residual of the Gaussian fit, but quite
strikingly, the mean stays at a value that is numerically very close to $3$,
with the result appearing to be exact as $n \to \infty$.  

This empirical finding warrants an explanation, which we provide subsequently
in Section \ref{sec:Theory}. In order to motivate our approach, we appeal to
progress in understanding the statistics of eigenvectors in the more general
setting of random matrix theory.  Recently, Bourgade, Huange, and Yau in
\cite{Bourgade:2017id} and Backhausz and Szegedy \cite{Backhausz:2016ot}
undertook a systematic study of eigenvector statistics of sparse random
matrices. They recovered specific information about the eigenvectors of
adjacency matrices of random regular graphs whose degree $z$ is bounded by the
number of vertices $n$.  Mirroring the hypothesis in
Ref.~\cite{Bourgade:2017id}, let $\delta>0$ be an arbitrarily small constant.
For a random $z$-regular graph which satisfies $n^\delta\le z\le
n^{2/3-\delta}$ their main result, Theorem 1.1, implies that
the entries of those eigenvectors of random regular graphs which are orthogonal
to the Perron-Frobenius mode $\wp$ have  asymptotically independent Gaussian
distributed entries. Moreover, it is well-known that vectors whose entries are
independently identically Gaussian distributed are uniformly distributed on the
unit sphere. (C.f. the textbook of Cram\'er \cite[Chapter 24]{Cramer:1999mms}
and the algorithmic implementation of this fact by Muller \cite{Muller:1959uds}). 
Together, these results suggest that the statistics of the inverse
participation ratio  can be directly investigated by computing moments of the
IPR function using the uniform probability distribution on a subsphere which is
orthogonal to $\wp$.

\section{Analysis of the IPR on a Hypersphere}
\label{sec:Theory}

In the previous section we numerically investigated the
distribution of IPR values for all non Perron-Frobenius eigenvectors of
each graph Laplacian across a large ensemble of graphs and empirically observed
that for $n \gg 1$:
\begin{equation}
    \expval{\frac{1}{n-1}\sum_{\vec{x} \in \mathcal{E}}
    \ipr(\vec{x})} \approx 3 \, . 
\label{eq:meanIPR}
\end{equation}
This result can be understood by exploiting the geometry of the set of
eigenvectors $\mathcal{E}=E\smsm \{\wp\}$ of each graph.  Their terminal points
lie on a hypersphere $S(\wp)$ which is orthogonal to the Perron-Frobenius mode
$\wp$, and is a subsphere of the standard real unit sphere
$S=\{\vec{u}\in\mathbb{R}^n \,:\ ||\vec{u}||=1\}$. In order to study the
properties of the inverse participation ratio on a space containing
$\mathcal{E}$,  we observe that the function $\ipr(\vec{x})$ is just a
polynomial of $n$ variables $x_i$ composed from the sum of fourth order
monomials $\ipr(\vec{x}) = n( x_1^4 + \cdots + x_n^4)$ which maps points
$\vec{x} \in S(\wp) \subset \mathbb{R}^n$ to $\mathbb{R}$.  As described above,
for random regular graphs with large $n$, the $x_i$ can be taken to be
independent and identically distributed according to the normal distribution
\cite{Bourgade:2017id, Backhausz:2016ot} and thus the vectors $\vec{x} \in
\mathcal{E}$ tend towards being uniformly distributed on $S(\wp)$. Thus in the
limit of large $n$, the eigenvector and graph averages in Eq.~(\ref{eq:meanIPR})
can be approximated by the continuous expectation value: 
\begin{equation}
    \mu^1_{\ipr} \equiv \expval{\ipr}_{S(\wp)} = \int_{S(\wp)} \ipr(\vec{x})
    P(\vec{x}) d\sigma (\wp)
\label{eq:aveIPRdef}
\end{equation}
where $d\sigma(\wp)$ is the measure and $P(\vec{x})$ the uniform distribution
on $S(\wp)$: 
\begin{equation}
    P(\vec{x}) = \frac{1}{\int_{S(\wp)}d \sigma(\wp)}\, .
    \label{eq:Px}
\end{equation}

Similarly the second moment, $\mu_{\ipr}^2$, corresponding to the variance of
the IPR distributions in Fig.~\ref{fig:rrdistIPR} can be investigated 
by averaging the eighth order polynomial $[\ipr(\vec{x})]^2$ on
$S(\wp)$ and employing the usual identity 
\begin{equation}
    \mu^2_{\ipr} \equiv \expval{\left(\ipr\right)^2}_{S(\wp)}
          - \left(\mu^1_{\ipr}\right)^2\ .
\label{eq:varIPRdef}
\end{equation}

Using Eq.~\eqref{eq:Px} in \eqref{eq:aveIPRdef} and \eqref{eq:varIPRdef} thus
allows us to compute the continuous first and second moments of the inverse
participation ratio  via the integration of fourth and eighth order polynomials
on $S(\wp)$.  This can be accomplished using the short note of Folland
\cite{Folland:2001tb} which provides a formula for integrating a polynomial
over a sphere.  Folland's formula is stated for a monomial $x^{\mbf{a}}=x_1^{a_1}\cdots
x_n^{a_n}$ and extends linearly to polynomials $\sum_{\mbf{a}} c_{\mbf{a}} \cdot x^{\mbf{a}}$ with
numerical coefficients $c_{\mbf{a}}$. Furthermore, the integral of a monomial is
dependent only on the Gamma function $\Gamma(b)=\int_0^\infty t^{b-1}e^{-t}\, dt$
where $b$ is a complex number with positive real part. To proceed, write
$d\sigma$ for the surface measure on the unit sphere $S\subset\mathbb{R}^n$.
Folland's result is the following. 

\begin{theorem}\cite{Folland:2001tb}\label{FollandsTheorem}
Let $x^{\mbf{a}}=x_1^{a_1}\cdots x_n^{a_n}$ be a monomial, 
so that $a_j \ge 0$ for all $1\le j\le n$. 
Setting $b_j=\frac{1}{2}(a_j+1)$, 
$$
\int_{S} x^{\mbf{a}}\ d\sigma
=\left\{ 
\begin{array}{cl}
0 & \textup{if some }a_j\textup{ is odd,} \\
& \\
\displaystyle\frac{2\Gamma(b_1)\Gamma(b_2)\cdots\Gamma(b_n)}
{\Gamma(b_1+b_2+\cdots+b_n)} 
& \textup{if all }a_j\textup{ are even.}
\end{array}
\right.
$$
\end{theorem}

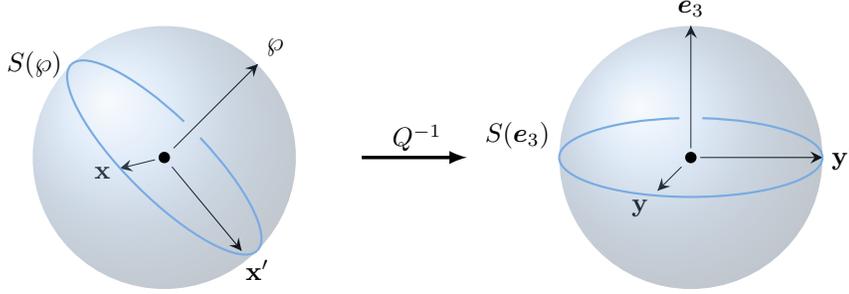
\begin{figure}[ht]
\begin{center}
\begin{tikzpicture}[scale=1.75]

\definecolor{green1}{rgb}{0.82,0.89,0.804}  
\definecolor{blue1}{rgb}{1.00, 0.34, 0.50}
\definecolor{blue1}{rgb}{1.00, 0.58, 0.47}
\definecolor{blue1}{rgb}{0.47,0.67,0.89}

\begin{scope}[shift={(-2,0)}, rotate=-45]
\node (A) at (0,0,0) {};
\node[above right] (B) at (0,1.0,0) {$\wp$};
\node[left] (C) at (-1,0,0) {$S(\wp)$};
\node[below] (D) at (0,0,0.86) {${\mathbf{x}}$};
\node[below] (E) at (1,0,0) {${\mathbf{x'}}$};
\draw[postaction={decorate,decoration={markings,mark=at position 1.0 with {\arrow[line width=1pt]{stealth}}}}](A)--(B);
\draw[postaction={decorate,decoration={markings,mark=at position 1.0 with
{\arrow[line width=1pt]{stealth}}}}](A)--(D);
\draw[postaction={decorate,decoration={markings,mark=at position 1.0 with
{\arrow[line width=1pt]{stealth}}}}](A)--(E);
\shade[ball color=blue1,opacity=0.3] (0,0,0) circle (1);
\draw[blue1, thick] (-1,0) arc (180:360:1cm and 0.3cm);
\draw[blue1, thick] (-1,0) arc (180:95:1cm and 0.3cm);
\draw[blue1, thick] (1,0) arc (0:85:1cm and 0.3cm);
\draw[black,thick, fill] (0,0,0) circle [radius=1pt] ;
\end{scope}

\draw[line width=0.3ex,->,style=-latex] (-0.5,0) -- (0.3,0) ;
\node at (-0.08,0.15) {$Q^{-1}$};

\begin{scope}[shift={(2,0)}]
\node (A) at (0,0,0) {};
\node[above] (B) at (0,1,0) {$\vec{e}_3$};
\node[above left] (C) at (-1,0,0) {$S(\vec{e}_3)$};
\node (D) at (0,0,1) {${\mathbf{y}}$};
\node[right] (E) at (1,0,0) {${\mathbf{y'}}$};
\draw[postaction={decorate,decoration={markings,mark=at position 1.0 with {\arrow[line width=1pt]{stealth}}}}](A)--(B);
\draw[postaction={decorate,decoration={markings,mark=at position 1.0 with {\arrow[line width=1pt]{stealth}}}}](A)--(D);
\draw[postaction={decorate,decoration={markings,mark=at position 1.0 with
{\arrow[line width=1pt]{stealth}}}}](A)--(E);
\shade[ball color=blue1,opacity=0.3] (0,0,0) circle (1);
\draw[blue1, thick] (-1,0) arc (180:360:1cm and 0.3cm);
\draw[blue1, thick] (-1,0) arc (180:95:1cm and 0.3cm);
\draw[blue1, thick] (1,0) arc (0:85:1cm and 0.3cm);
\draw[black,thick, fill] (0,0,0) circle [radius=1pt] ;
\end{scope}
\end{tikzpicture}
\end{center}
\caption{\label{fig:SphereRotation}The rotation procedure for
$n=3$ from an oriented to standard subsphere (circle) embedded in $\mathbb{R}^3$.}
\end{figure}

We wish to average the polynomial $\ipr(\vec{x})$ over the 
subsphere $S(\wp) \subset S$ that is orthogonal to the Perron-Frobenius vector
$\wp$.  Thus, in order to use Theorem~\ref{FollandsTheorem} we must first
rotate our subsphere $S(\wp)$ around a $(n-2)$-dimensional subspace to coincide
with the subsphere $S({\vec{e}}_n)$ as depicted for the case $n=3$ in
Fig.~\ref{fig:SphereRotation}.  
%
%

Note that although the target sphere $S({\vec{e}}_n)$ is defined by coordinates
in $\mathbb{R}^n$, the $n^\mathrm{th}$ coordinate of each of its points equals
zero. Hence, $S({\vec{e}}_n)$ is realized as the unit sphere of
$\mathbb{R}^{n-1}$ which is embedded in $\mathbb{R}^n$ according to the rule
$(y_1,\ldots, y_{n-1})\mapsto (y_1,\ldots, y_{n-1},0)$. Changing variables
therefore allows the direct application of Theorem~\ref{FollandsTheorem} to the
unit sphere in $\mathbb{R}^{n-1}$ to achieve our result. The remainder of this
section is devoted to these analytic calculations. 


\subsection{The rotation matrix $Q$}

The required change of variables is performed via a rotation matrix $Q \in
\mathrm{SO}(n)$ which has the property that for any $\vec{y} \in S({\vec{e}_n})$ there
exists $\vec{x} \in S(\wp)$ such that 
\begin{equation}
    \vec{y} = Q \vec{x}.
\label{eq:Qdef}
\end{equation}
We begin the construction of $Q$ by using the Gram-Schmidt process to find two orthonormal
vectors in the plane defined by ${\vec{e}}_n$ and $\wp$: 
\begin{align}
    \vec{v}_1 &= \wp = \frac{(1,\ldots,1)}{\sqrt{n}} \\
    \vec{v}_2 &= \frac{\vec{e}_n - \left(\wp \cdot \vec{e_n}\right)\vec{e}_n}
    {||\vec{e}_n - \left(\wp \cdot \vec{e_n}\right)\vec{e}_n||} 
    = -\frac{(1,\ldots,1,1-n)}{\sqrt{n(n-1)}}.
\end{align}
To align $\wp$ with $\vec{e}_n$, we need to perform a rotation by an angle
$\theta$ defined by: 
\begin{equation}
    \cos \theta = \wp \cdot \vec{e}_n = \frac{1}{\sqrt{n}}
\label{eq:theta}
\end{equation}
around the plane formed by $\vec{v}_1$ and $\vec{v}_2$ and the identity space
spanned by the $(n-2)$-dimensional complement of the orthonormal basis. 
Hence, we have 
\begin{equation}
Q = \mathbbm{1} + \sin \theta \left( \vec{v}_2 \otimes \vec{v}_1 - \vec{v}_1 \otimes \vec{v}_2 \right)
+ (\cos \theta - 1) \left( \vec{v}_1 \otimes \vec{v}_1 + \vec{v}_2 \otimes
\vec{v}_2\right)
\label{eq:Qmatrix}
\end{equation}
where $\mathbbm{1}$ is the $n\times n$ identity matrix and $\otimes$ represents
the tensor product of vectors.  Considering vector components: $v_{1i} =
1/\sqrt{n}$ and $v_{2i} = (n\delta_{ni}-1)/\sqrt{n(n-1)}$ we express the
rotation matrix in a form more useful for performing explicit calculations:
\begin{equation}
    Q_{ij} = \delta_{ij} +
    \frac{1}{\sqrt{n}}\left[\frac{1-\sqrt{n}}{n-1}\left(1-\delta_{in} -
    \delta_{nj} + n\delta_{nj}\delta_{in}\right) + \delta_{in} -\delta_{nj}
\right].
\label{eq:Qcomponent} 
\end{equation}
Using Eq.~(\ref{eq:Qcomponent}) it is therefore straightforward to confirm that:
\begin{align*}
    (i) \,\, & \sum_{j=1}^n Q_{ij} = \sqrt{n} \delta_{in},\,\, \text{and} \\
    (ii) \,\, & \forall\ \vec{x} \in \mathbb{R}^n\; \text{such that}\; || \vec{x} || = 1 \; \text{and}\;
    \vec{x} \cdot \wp = 0,\; y_n = \sum_{j=1}^{n} Q_{nj} x_j = 0\; .
\end{align*}

\subsection{Evaluation of the inverse participation ratio moments}

Folland's straightforward application of Theorem~\ref{FollandsTheorem} shows
that the $(n-2)$-dimensional measure of each of the spheres $S(\vec{e}_n)$ and
$S(\wp)$ is
$2\pi^{(n-1)/2} / \Gamma(\frac{n-1}{2})$, 
whose reciprocal is $P(\vec{x})$, the probability of uniformly choosing a point from such a
sphere. To proceed, write $d\sigma(\vec{e}_n)$ for the $(n-2)$-dimensional
surface measure of the sphere $S(\vec{e}_n)$ and $d\sigma(\wp)$ for the
$(n-2)$-dimensional surface measure of the sphere $S(\wp)$.  
Applying the multivariable change of basis formula for integrals to
Eq.~\eqref{eq:aveIPRdef} therefore yields 
\begin{align}
    \expval{\ipr}_{S(\wp)}
& =\int_{S(\wp)} \ipr(\vec{x}) \
\frac{\Gamfunc{n-1}{2}}{2\pi^{(n-1)/2}}  d\sigma(\wp) \nonumber
\\
&=\frac{\Gamfunc{n-1}{2}}{2\pi^{(n-1)/2}}  \ \int_{S(\vec{e}_n)}
\ipr\left(Q^{-1}\vec{y}\right) \left |J\left(Q^{-1}\right) \right| d\sigma(\vec{e}_n). 
\label{eq:Qtransform}
\end{align}
Note that $Q$ is an orthonormal rotation matrix, so that the Jacobian
$\lvert J(Q^{-1})\rvert=1$. 
As mentioned above, $y_n=0$ since $\vec{y}\cdot\vec{e}_n=0$, so that all our
monomials have $(n-1)$ variables.  Theorem \ref{FollandsTheorem} further
guarantees that although the expansions of $\ipr(Q^{-1}\vec{y})$ and
$[\ipr(Q^{-1}\vec{y})]^2$ have many monomial terms with odd exponents, our
moment calculations give non-zero values for only those monomial terms having
all their exponents even.  Combining the structure of $Q^{-1}=Q^\intercal$, the
multinomial theorem, the results of averaging relevant monomials $\langle
y_k^{a_k} \cdots y_{k'}^{a_{k'}}\rangle_{S(\vec{e}_n)}$ over the sphere 
found in \ref{sec:Gamma}, and the expansions in \ref{sec:Qsums} we now give 
closed forms for the first and second moments of the inverse participation
ratio over $S(\wp)$.  
%
%
Applying Eq.~\eqref{eq:Qtransform}, the first moment is 
\begin{align}
    \expval{\ipr}_{S(\wp)}
&= \frac{\Gamfunc{n-1}{2}}{2\pi^{(n-1)/2}}  \int_{S(\vec{e}_n)}
\ipr\left(Q^{-1}\vec{y}\right)\ d\sigma(\vec{e}_n) \nonumber \\
&= n \sum_{i=1}^n\left( 
\sum_{k=1}^{n-1}\expval{y_k^4}_{S(\vec{e}_n)}Q_{ki}^{4}
    + 3 \sideset{}{'}\sum_{k,\ell}^{n-1} \expval{y_k^2 y_\ell^2}_{S(\vec{e}_n)} 
Q_{ki}^{2}Q_{\ell i}^{2}\right) \nonumber \\
&= \frac{3n}{4}\frac{\Gamfunc{n-1}{2}}{\Gamfunc{n+3}{2}}
\sum_{i=1}^n\left( \sum_{k=1}^{n-1} Q_{ki}^{4}
+ \sideset{}{'}\sum_{k,\ell}^{n-1}  Q_{ki}^{2}Q_{\ell i}^{2}\right) 
\label{eq:mu1}
\end{align}
where we have used the notation that a prime on a multiply indexed sum
enforces the constraint that no equal indices are included, i.e.
\begin{equation*}
    \sideset{}{'}\sum_{k,\ell}^{n-1} (\cdots)  \equiv \sum_{k\ne \ell}^{n-1}
    (\cdots) \equiv \sum_{k=1}^{n-1}
    \sum_{\ell=1}^{n-1} (\cdots) \left(1-\delta_{k \ell}\right)
\end{equation*}
and the monomial averages over the subsphere $\expval{y^4}_{S(\vec{e}_n)}$ and
$\expval{y^2y^2}_{S(\vec{e}_n)}$ have been computed using Theorem~\ref{FollandsTheorem} with
the individual results given in Table~\ref{tab:Fours}. The double and triple
summations over the components of the rotation matrix $Q$ are evaluated in
\ref{sec:Qsums} and substituting Eqs.~\eqref{eq:Q41} and \eqref{eq:Q221} into
Eq.~\eqref{eq:mu1} yields:
\begin{align}
    \mu_{\ipr}^1
&= \frac{3n}{(n+1)(n-1)} \left[n - \frac{29 + 30\sqrt{n}+5n}{(1+\sqrt{n})^2}
+\frac{24}{\sqrt{n}} - \frac{9}{n} \right.  \nonumber \\
& \left. \qquad \qquad \qquad \quad +\; 
\frac{(\sqrt{n}-1)(3\sqrt{n}+5)(n-2)}{n(\sqrt{n}+1)^2}\right] \nonumber \\
\label{eq:mu1IPR}
& = 3 - \frac{6}{n+1}
\end{align}
which has the observed limiting value of $3$ for $n\gg 1$.

%
The calculation of the second central moment proceeds in a similar fashion by
using Eq.~\eqref{eq:varIPRdef} and applying the above general preliminaries to
the square of the inverse participation ratio polynomial. We have 
\begin{align}
    & \expval{\ipr^2}_{S(\wp)}  \nonumber \\
&\;  = \frac{\Gamfunc{n-1}{2}}{2\pi^{(n-1)/2}}
    \int_{S(\vec{e}_n)}
    \left[\ipr\left(Q^{-1}\vec{y}\right)\right]^2\  \ d\sigma(\vec{e}_n) \nonumber \\
    &\; =
n^2\sum_{k=1}^{n-1}\expval{y_k^8}_{S(\vec{e}_n)}\left[\sum_{i=1}^{n}Q_{ki}^{8} +
   \sideset{}{'}\sum_{i,j}^n
        Q_{ki}^{4}Q_{kj}^{4} \right]\nonumber  \\
    &\;\;\; + n^2\sideset{}{'}\sum_{k,\ell}^{n-1} \expval{y_k^6 y_\ell^2}_{S(\vec{e}_n)}\left[28 \sum_{i=1}^{n}
    Q_{ki}^{6}Q_{\ell i}^2 + \sideset{}{'}\sum_{i,j}^n \left(12
            Q_{ki}^{4}Q_{\ell j}^{2}Q_{kj}^{2} +
16Q_{ki}^{3}Q_{li}Q_{kj}^{3}Q_{\ell j} \right)\right] \nonumber  \\
&\;\;\;+ n^2\sideset{}{'}\sum_{k,\ell}^{n-1}\expval{y_k^4 y_\ell^4}_{S(\vec{e}_n)} 
\left[35 \sum_{i=1}^{n} Q_{ki}^{4}Q_{\ell i}^4
+ \sideset{}{'}\sum_{i,j}^n \left( Q_{ki}^{4}Q_{\ell j}^{4}
    + 18 Q_{ki}^{2}Q_{\ell i}^{2}Q_{kj}^{2}Q_{\ell j}^{2} \right. \right. \nonumber \\
& \qquad \qquad \qquad \qquad \;\;\; 
\left. \phantom{\sum_k^n} \left. +\; 16 Q_{ki}^{3}Q_{\ell i}Q_{kj}Q_{\ell j}^{3} \right) \right]\nonumber  \\
&\;\;\;+ n^2\sideset{}{'}\sum_{k,\ell,m}^{n-1} \expval{y_k^4 y_\ell^2y_m^2}_{S(\vec{e}_n)}
\left[210 \sum_{i=1}^{n} Q_{ki}^{3}Q_{\ell i}^3Q_{m i}^3 +
\sideset{}{'}\sum_{i,j}^n \left( 6Q_{ki}^{4}Q_{\ell j}^{2}Q_{mj}^2
    \right. \right.\nonumber  \\
& \qquad \qquad \qquad \qquad \qquad   \left. \vphantom{\sum_1^2} \left.   
    +\; 72 Q_{ki}^{2}Q_{\ell i}Q_{mi}Q_{k j}^{2}Q_{\ell j}Q_{mj} 
+ 96 Q_{ki}^{3}Q_{\ell i}Q_{kj}Q_{\ell j}Q_{mj}^{2} \right) \right]\nonumber  \\
&\;\;\;+ n^2\sideset{}{'}\sum_{k,\ell,m,p}^{n-1} \expval{y_k^2 y_\ell^2y_m^2 y_p^2}_{S(\vec{e}_n)} 
\left[105 \sum_{i=1}^{n} Q_{ki}^{2}Q_{\ell i}^2Q_{m i}^2 Q_{p i}^2
+ \sideset{}{'}\sum_{i,j}^n \left( 9Q_{ki}^{2}Q_{\ell i}^2Q_{mj}^2Q_{p j}^{2}
    \right.  \right.\nonumber  \\
\label{eq:mu2sum}
&\qquad\; \qquad \qquad  \left. \vphantom{\sum_1^2} \left. 
+\;  72 Q_{ki}^{2}Q_{\ell i}Q_{mi}Q_{p j}^{2}Q_{\ell j}Q_{mj} + 
24 Q_{ki}Q_{\ell i}Q_{mi}Q_{pi}Q_{kj}Q_{\ell j}Q_{mj}Q_{pj} \right)
\right]
\end{align}
and combining the results of Table~\ref{tab:Eights} and \ref{sec:Qsums} we find
\begin{equation}
    \expval{\ipr^2}_{S(\wp)}
= 9 + \frac{48}{n+1} - \frac{270}{n+3} + \frac{210}{n+5}.
\label{eq:mu2}
\end{equation}
Subtracting the square of the first moment, we arrive at the final
expression for the second moment of the inverse participation ratio on
$S(\wp)$

\begin{equation}
    \mu_{\ipr}^2 = \frac{24 n(n-2)(n-3)}{(n+5)(n+3)(n+1)^2}
\label{eq:mu2IPR}
\end{equation}
which indeed tends to zero as $n \to \infty$.

\section{Comparison With Exact Diagonalization Results}
\label{sec:Experiments}

Having uncovered the origin of the universal number $3$ as
$\lim_{n\to\infty} \mu^1_{\ipr}$ for the mean of the continuous 
IPR, we now undertake a systematic comparison of exact diagonalization results
for the inverse participation ratio of the Laplacian on finite sized random
regular graphs and the predictions on subspheres embedded
in $\mathbb{R}^n$ as a function of $z$ and $n$. 

\subsection{1st IPR moment}
\label{sub:1st_ipr_moment}

\begin{figure}[ht]
\begin{center}
\includegraphics[width=0.75\columnwidth]{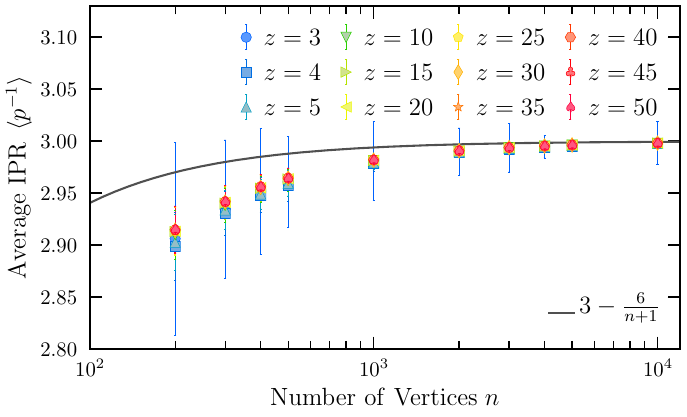}
\end{center}
\caption{Graph and mode averaged inverse participation ratio vs. the number of
vertices $n$ for different graph degrees $z$ (symbols). The solid line
shows the finite size prediction for $\mu^1_{\ipr}$ obtained by averaging over
a sphere.}
\label{fig:rraveIPRvsn}
\end{figure}

The finite size scaling behavior of the first moment of the IPR can be
quantified by explicitly computing the average of the IPR over all
non-Perron-Frobenius eigenvectors $\vec{x} \in \mathcal{E}$, and then further
averaging this quantity over graph realizations.  In particular, we define the
mode-averaged IPR (first IPR moment) for a given graph to be
\begin{equation}
    {p^{-1}}= \frac{1}{n-1} \sum_{\vec{x} \in \mathcal{E}} 
    \ipr(\vec{x}),
\label{eq:aveIPR}
\end{equation}
while $\expval{p^{-1}}$ includes an additional average over the graph ensemble.
Fig.~\ref{fig:rraveIPRvsn} depicts the $n$ dependence of this quantity 
for all graph degrees considered, where we have averaged over $N_G = 5000$
random regular graphs for $n < 5000$ and $N_G=1000$ graphs for $n \ge 5000$.
%
%

The solid line describes the function $\mu^1_{\ipr} = 3-6/(n+1)$ derived in
Eq.~\eqref{eq:mu1IPR} by averaging the IPR polynomial over the subsphere 
$S(\wp)$. There is good agreement for $n > 1000$, seemingly independent of
graph degree.  The error bars are obtained by computing the standard deviation
of ${p^{-1}}$ over all graphs in the generated set, with the largest
uncertainties occurring for $z=3$.  We postpone a discussion of the size and
$z$-dependence of graph-to-graph fluctuations until the end of this section.

\begin{figure}[ht]
\begin{center}
\includegraphics[width=0.495\columnwidth]{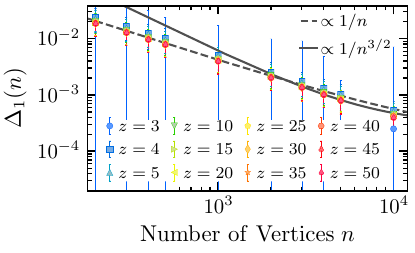}
\includegraphics[width=0.495\columnwidth]{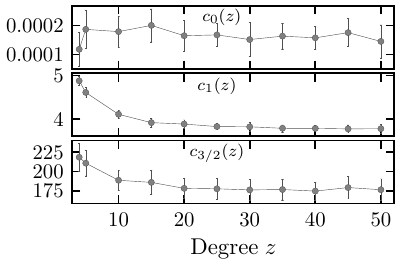}
\end{center}
\caption{Left: Normalized deviation of the graph and mode averaged inverse
    participation ratio for the Laplacian on random regular graphs from the
    sphere averaged value of $\mu^1_\ipr$, given in Eq.~(\ref{eq:Delta1}) in
    the text. Two fits to the residual data corresponding to $c_0 +
    c_1/n$ (dashed) and  $c_{3/2}/n^{3/2}$ (solid) are
    shown with the latter only using data with $n > 1000$. Right:  
$z$-dependent fitting parameters are consistent with vanishing $O(1)$
corrections to to $\mu^1_{\ipr}$.}
\label{fig:rraveIPRvsz}
\end{figure}

We may investigate deviations between $\expval{p^{-1}}$ and $\mu_{\ipr}^1$ 
by defining a normalized residual
\begin{equation}
    \Delta_1(z,n) = 1 - \frac{\expval{p^{-1}}}{\mu^1_\ipr}
\label{eq:Delta1}
\end{equation}
which is plotted in Fig.~\ref{fig:rraveIPRvsz} (left) as a function of $n$ for
different values of $z$.
%
%
The residual decays with increasing $z$ and $n$ with the correction being fit
by an empirically determined function of the form $c_0(z) + c_1(z)/n$ (dashed
line) where the extracted coefficients are shown in the right panel of
Fig.~\ref{fig:rraveIPRvsz}.  The exact diagonalization data is consistent with
the absence of any $O(1)$ correction to Eq.~(\ref{eq:mu1IPR}) for large $z$
within errorbars, i.e.~$c_0(z\to\infty)\to0$. The coefficient $c_1(z)$ appears to decay only
weakly with increasing $z$ and a $z$-dependent $1/n$ correction cannot be ruled
out at the level of our statistical uncertainty for $n \le 1000$. For $n \ge
1000$ the residual can also be described by a function of the form
$c_{3/2}(z)/n^{3/2}$ as shown by the solid line in Fig.~\ref{fig:rraveIPRvsz}
(left) supporting the leading finite $n$ behavior of $\mu^1_{\ipr}$.

\subsection{2nd IPR moment}
\label{sub:2nd_ipr_moment}

\begin{figure}[ht]
\begin{center}
\includegraphics[width=0.75\columnwidth]{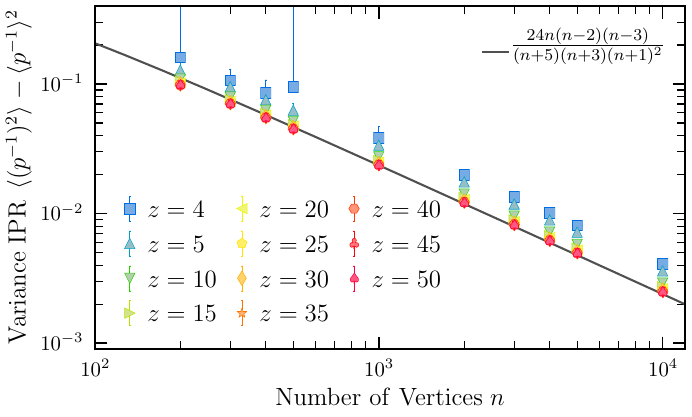}
\end{center}
\caption{The graph averaged second moment of the mode averaged inverse
    participation ratio distribution for random regular graphs of varying
    degree vs. the number of vertices. The solid line shows the sphere averaged
value, $\mu^2_{\ipr}$.}
\label{fig:rrvarIPRvsn}
\end{figure}

Next, we consider the prediction of Eq.~\eqref{eq:mu2IPR} by studying the
second moment of the distribution of IPR values on finite sized random regular
graphs: $\expval{{(p^{-1})^2}} - \expval{p^{-1}}^2$  averaged over $N_G =
5000$ unique graphs for $n < 5000$ and $N_G = 1000$ for $n \ge 5000$.  The
results are shown in Fig~\ref{fig:rrvarIPRvsn}, where now deviations from the
sphere-averaged value $\mu^2_{\ipr}$ (included as a solid line) are observed for
all values of $n$ and $z$ considered. 
%
%

The degree dependence is the most obvious: the exact diagonalization results
are systematically larger than $\mu^2_{\ipr}$ for small z, with the discrepancy
decreasing as $z$ increases. We have not included data for $z=3$ in
Fig.~\ref{fig:rrvarIPRvsn} as these points lie mostly off the scale and the
peculiarities of this degree will be carefully investigated in the following
subsection.  Additionally, due to the logarithmic scale, we have only plotted
errorbars showing the additive uncertainty across graphs.  For $z>4$, the
standard deviation is on the order of the symbol sizes. 

\begin{figure}[ht]
\begin{center}
\includegraphics[width=0.495\columnwidth]{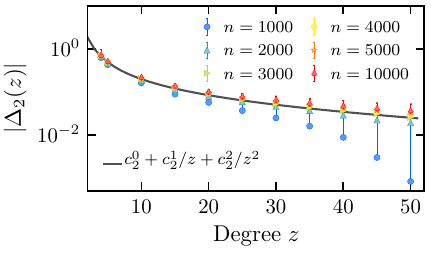}
\includegraphics[width=0.495\columnwidth]{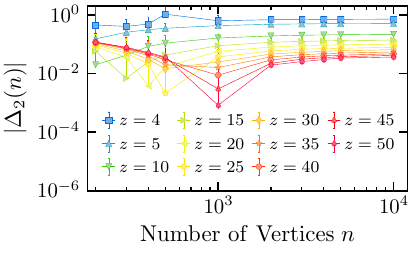}
\end{center}
\caption{The normalized positive residual between the second moment of the inverse
participation ratio distribution for large random regular graphs and the value
obtained by averaging over a sphere.  The left panel shows that the $z$
dependence of the correction can be fit via a second order polynomial in $1/z$
with an offset that persists in the $n\to\infty$ limit (right panel).}
\label{fig:rrerrIPRvsz}
\end{figure}

We again define a normalized residual for the second moment:
\begin{equation}
    \Delta_2(z,n) = 1 - \frac{\expval{(p^{-1})^2} - 
    \expval{p^{-1}}^2} {\mu^2_\ipr} 
\label{eq:Delta2}
\end{equation}
and the absolute value $|\Delta_2(z,n)|$ is plotted in Fig.~\ref{fig:rrerrIPRvsz}.
%
%

Here, the dominant deviations from the first sphere averaged value
$\mu_{\ipr}^1$ are degree dependent, and they can be described by a function of
the form $c_2^0 + c_2^1/z + c_2^2/z^2$.  The values of the fitting constants
$c_2^k$ depend on $n$, with the solid line in the left panel representing their
average values for $n > 1000$.  Different fitting functions were investigated,
including those with non-integer negative powers of $z$, but the resulting
second order polynomial in $1/z$ provided the optimal value of the least square
fitting $\chi^2$ value. The right panel of Fig.~\ref{fig:rrerrIPRvsz} shows the
$n$ dependence of $\Delta_2$ and it appears that it remains non-zero even as
$n\to\infty$.  This is consistent with the fitting parameter $c^2_0$ which is
finite within errorbars for all values of $z$ considered.

\subsection{Effects of localized eigenvectors}
\label{subsec:_localized_eigenvectors}
\begin{figure}[ht]
\begin{center}
\includegraphics[width=0.495\columnwidth]{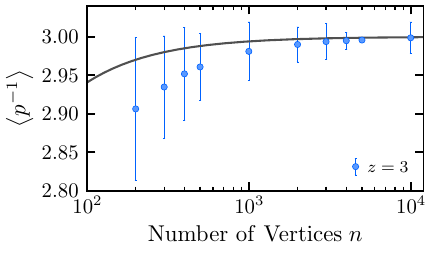}
\includegraphics[width=0.495\columnwidth]{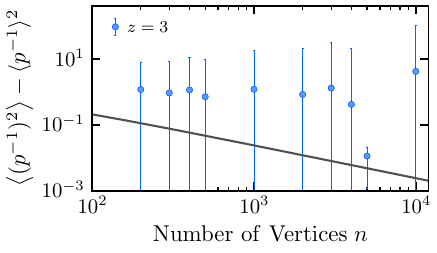}
\end{center}
\caption{The first (left) and second (right) moments of the inverse
    participation ratio distribution vs. the number of vertices computed via
    exact diagonalization of $5000$ random regular graphs with degree $z=3$.
    The errorbars correspond to one standard deviation, and they are
    significantly larger than those observed for $z>3$. Solid gray lines are
the predicted sphere averaged values of $\mu_{\ipr}^1$ and $\mu_{\ipr}^2$
defined in Eqs.~\eqref{eq:mu1IPR} and \eqref{eq:mu2IPR} respectively.}
\label{fig:zeq3}
\end{figure}

We now address the issue of the large graph-to-graph variance observed around
the first and second moments of the IPR distribution for small $z$.
This data is displayed in Fig.~\ref{fig:zeq3} where the graph averaged first
and second moments of the IPR distribution are shown as a function of $n$ for
$z=3$.  
%
%
The displayed errorbars correspond to one standard deviation and we observe
that the effects are most pronounced for the variance of the IPR. Data
points consistently fall above the sphere averaged value of $\mu_{\ipr}^2$, and
the mean value between graphs can vary by as much as $1000\%$.  However, for
both the first and second moments, the data is consistent with values of
$\mu^1_{\ipr}$ and $\mu^2_{\ipr}$ computed using integration.  The
existence of a single outlying data point corresponding to $n=5000$ that is
much closer to $\mu^2_{\ipr}$ in the right panel of Fig.~\ref{fig:zeq3} is
suggestive that the sample set of unique random regular graphs may not be large
enough to capture the variation in the eigenvector components amongst graphs as
measured by the inverse participation ratio.

\begin{figure}[ht]
\begin{center}
\includegraphics[width=0.495\columnwidth]{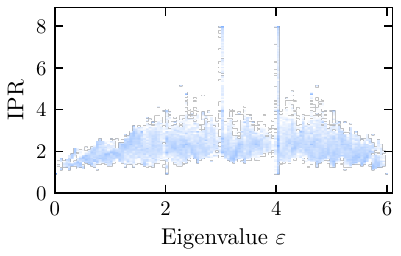}
\includegraphics[width=0.495\columnwidth]{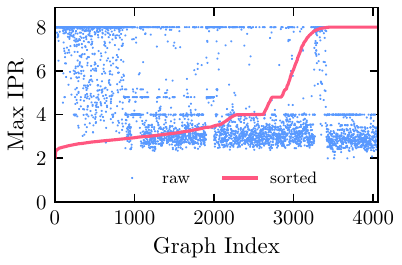}
\end{center}
\caption{A histogram with shading on a logarithmic scale of the inverse participation
ratio vs. eigenvalue for all 4060 random regular graphs with $n=16$ and
$z=3$ (left). The right panel shows the maximum value of the inverse
participation ratio across all modes for a given graph plotted as a function of
a fictitious graph index.  By relabeling the graph index we also include the
maximum value sorted by size.}
\label{fig:z3n16}
\end{figure}

To better understand the prevalence of this effect for graphs of small degree,
we have exactly diagonalized the Laplacian for every one of the $\mathcal{N}_G =
4060$ unique random regular graphs with $n=16$ and $z=3$ \cite{Meringer:1999kx}.
Analyzing the eigenvectors and computing the inverse participation ratio, we
find:  
\begin{align}
    \expval{p^{-1}} & = 2.4 \pm  0.4 \\
     \expval{(p^{-1})^2} - 
\expval{p^{-1}}^2 &= 0.95  \pm  1.2\; .
\end{align}
The origin of these sizeable graph-to-graph variations is uncovered in the left
panel of Fig.~\ref{fig:z3n16}, which shows a histogram of all IPR values
(excluding the Perron-Frobenius mode) for the complete graph set plotted
against their corresponding eigenvalue.
%
%
The frequency of IPR values is shown on a logarithmic color scale from light to
dark, and we observe two spikes near $\varepsilon=z$ and $\varepsilon = z+1$
with the inverse participation ratio ranging up to its maximal value of $n/2 =
8$.  In the right panel of Fig.~\ref{fig:z3n16} we show the maximum value of
the IPR across all $n=16$ modes and find at least one value of $8$ for nearly
$15\%$ of graphs in the set.  

These graph realizations contain special eigenvectors $\{\vec{v}\}
\subset \mathcal{E}$ of the
Laplacian, with $\ipr(\vec{{v}}) = n/2$.  More generally, a vector with
exactly $k$ equal non-zero sites is of the form
\begin{equation} 
    v_i = \frac{(-1)^{q_i}}{\sqrt{k}} \delta_{i,i(k)}
\end{equation}
where $i(k) \in \{i_1,i_2,\cdots,i_{k-1},i_k\}$ and $q_i \in\{0,1\}$ such that
$\sum_{i(k)}(-1)^{q_i} =0$.  Clearly this is only possible for $k = 2m$ where
$m \in \{1,\cdots,\lfloor n /2 \rfloor\}$, with the total number of such vectors 
given by the multinomial coefficient
\begin{equation*}
    {n\choose{m,\ m,\ n-2m}}=\frac{n!}{m!\ m!\ (n-2m)!}\; . 
\end{equation*}
For vectors of this form, the inverse participation ratio is given by
\begin{equation} 
    \ipr(\vec{{v}}) = n \sum_{i=1}^n |{v}_i|^4 = \frac{n}{k^2} k =
\frac{n}{k} \end{equation}
which is exactly what we observe for $k = 2$. 
This maximal value for the IPR occurs for the most localized mode that is still compatible with orthonormality
to the Perron-Frobenius eigenvector and consists of exactly two non-zero values
with opposite sign.  We have confirmed that such eigenvectors indeed appear 
in our large-$n$ graph ensembles for $z=3$. 
Such vectors have a maximal nodal domain count of unity
\cite{Dekel:2010dz} and we observe that almost all eigenvectors with
$\ipr(\vec{v}) = n/2$ have non-zero components of opposite sign situated 
in vector components with consecutive indices. 

The large graph-to-graph variations displayed in Fig.~\ref{fig:zeq3} can thus
be traced back to these localized eigenvectors in combination with the
prefactor of $n$ in the definition of the IPR given in Eq.~(\ref{eq:ipr}).  By
averaging over the sphere, we found in Eq.~\eqref{eq:mu1IPR} that
$\mu_{\ipr}^1 \sim 3$ for $n\gg1$.  However as just demonstrated, localized
eigenvectors can contribute IPR values of $\mathrm{O}(n)$ to the first moment.
This implies that the variance around the mean could contain dominant terms
scaling like $n^2$ which will always have an effect when averaging over a
finite number of large graphs. 

\section{Discussion}
\label{sec:Discussion}

In this paper we have investigated the first and second moments of the
distribution  of the inverse participation ratio for all eigenvectors of the
discrete Laplacian on finite size random regular graphs. 
By exactly diagonalizing large ensembles of graphs of up to $n=10000$ vertices
we find that the first moment of the inverse participation ratio approaches a
constant of order unity  $\lim_{n\to\infty} \expval{p^{-1}} =
3$ for all values of $z$. This result can be understood in terms of an
analytically determined value for the average inverse participation ratio
$\mu_{\ipr}^1 = 3 - 6/(n+1)$ obtained by averaging a fourth order
polynomial corresponding to the IPR over the sphere $S(\wp)$ with uniform
probability measure.  We take this agreement as additional evidence that that the
\emph{average} eigenvector of the Laplacian on random regular graphs is
delocalized, with its components tending towards being independent and
identically distributed Gaussian random variables.  For smaller values of $n$
that do not necessarily satisfy the constraint $n^\delta\le z\le
n^{2/3-\delta}$ for $\delta > 0$ \cite{Bourgade:2017id}, we observe deviations
from $\mu_{\ipr}^1$ at $O(1/n)$ that could be potentially useful when
quantifying the distance from uniformity for a given set of random regular
graph eigenvectors.  The methodology used here to average the IPR
polynomial over the hypersphere with constant probability could be employed to
study other observables of physical interest on random graphs when $n$ is large.

For the variance of the inverse participation ratio computed over all modes, we
have again compared our exact random regular graph eigenvectors with an
analytical result from continuous averaging over $S(\wp)$ with uniform measure
where we find  $\mu_{\ipr}^2 = 24 n(n-2)(n-3)/[(n+5)(n+3)(n+1)^2]$. Here we
observe weaker agreement that now strongly depends on the graph degree.  This
discrepancy appears to persist even in the limits $z,n\to\infty$.  When
computing the standard deviation of the IPR over an ensemble of up to $5000$
random regular graphs, we find that for small values of the graph degree $z$,
large fluctuations between graph eigenvectors can cause variations in the first
and second moments as large as $1000\%$. By analyzing the complete set of
graphs for $z=3$ and $n=16$ we have shown that such deviations may arise from
graphs where the Laplacian has localized eigenvectors consisting of only a few
non-zero elements and thus IPR values of $n/2$.  Although we have no proof that
these vectors appear as eigenvectors of the Laplacian for finite size random
regular graphs in general, we have demonstrated that there are factorially many
such eigenvectors that are orthonormal to the Perron-Frobenius mode.

In general, the large but finite ensemble size of random regular graphs we
analyze is much smaller than the total number of random regular graphs, 
which is known \cite{Bollobas:1980} to asymptotically grow exponentially with $n$.
Hence, the fact that the standard deviation in the mean and
variance of the inverse participation ratio for large $n$ and $z$ appear to be
small is likely due to our samples of random regular graphs not being fully
representative of the eigenvector variation which exists. 

As $z$ and $n$ increase, extremely large ensembles
of graphs need to be studied in order to to balance the dominant effects of
localized modes, especially for non-linear observables. Thus, any observed
deviation of the 2nd moment of the IPR from its uniform value for a given
set of finite size graphs could be employed as a proxy for the representative
suitability of the sampled set when $n$ is large.  This may may have practical implications for
studying physical models with observables computed on regular graphs.

It would be interesting to explore this issue further, although considerable
computational resources would have to be employed to diagonalize large
numbers of graph Laplacians for $n \gg 1$.  Determining the combinatorial, physical, and
theoretical significance of these localized eigenvectors is thus left
as a topic of future work. 

\section{Acknowledgements}
\label{sec:acknowledgements}
The authors thank R. Bauerschmidt for help clarifying current results on the
distribution of eigenvector components of random regular graphs. A.D. is
deeply indebted to Z.~Te\v{s}anovi\'{c}, for diverse and stimulating
discussions that ultimately spawned my interest in graph theory and lead to
this collaborative work. A.D. also thanks T. Lakoba for his insights into high
dimensional coordinate transformations.  T.C. is grateful to M.~Shah, for
lively and informative numerical linear algebra conversations and to B. H.~Lee 
for clarification of key statistical notions.
This research was supported in part by the National Science Foundation (NSF)
under award No.~DMR-1553991 (A.D.). All computations were performed on the
Vermont Advanced Computing Core supported in part by NSF award No.~OAC-1827314.
\\

\appendix

\section{Averages of Monomials on the Sphere}
\label{sec:Gamma}
This appendix contains results of averaging monomials over the uniform
distribution on $S(\vec{e}_n)$ necessary for the integral calculations of
Section \ref{sec:Theory}. Note that in our case, the polynomial
$\ipr(Q^{-1}\vec{y})$ consists of monomials in the variables $y_1, \ldots,
y_{n-1}$ since $y_n=0$. Theorem \ref{FollandsTheorem} guarantees that the
integral of a monomial is non-zero precisely when its variables have even
exponents, and as such, we give values in only this case. 

When performing the first moment calculations we apply
Theorem~\ref{FollandsTheorem} to two distinct monomial types. The monomial
$y_i^4$ has $a_i=4$ for a fixed $i$ and $a_j=0$ for all $j\ne i$, while the
monomial $y_i^2y_j^2$ has $a_i=a_j=2$ and $a_k=0$ for $k\ne i,j$.  In the first
case, we therefore have $b_i=5/2$ for a single $i$ and $b_j=1/2$ for $j\ne i$.
In the second case, $b_i=b_j=3/2$ and $b_k=1/2$ for $k\ne i,j$.  For $y_k^4$ we
average over $S(\vec{e}_n)$ and find:
\begin{align}
    \expval{y_k^4}_{S(\vec{e}_n)} &= \int_{S(\vec{e}_n)} y_k^4\, P(\vec{y})\, 
    d\sigma(\vec{e}_n) \nonumber \\
    &= \frac{\Gamfunc{n-1}{2}}{2 \pi^{(n-1)/2}}
    \frac{2\Gamfunc{5}{2}\left[\Gamfunc{1}{2}\right]^{n-2}}
    {\Gamfunc{n+3}{2}} \nonumber \\
    &=
    \frac{3}{4}\frac{\Gamfunc{n-1}{2}}{\Gamfunc{n+3}{2}}
\label{eq:y4ave}
\end{align}
This result along with the similarly computed $\expval{y^2 y^2}$ term are
gathered in Table~\ref{tab:Fours}.
\begin{table}[ht]
\begin{center}
\renewcommand{\arraystretch}{1.5}
\setlength\tabcolsep{4pt}
\begin{tabular}{@{}p{4em}cc} 
\toprule
\textbf{Average:} & 
$\expval{y_k^4}_{S(\vec{e}_n)}$ & 
$\expval{y_k^2y_\ell^2}_{S(\vec{e}_n)}$\\
\addlinespace[0.4em]
\midrule
\addlinespace[0.5em]
\textbf{Value:}
& $\ds\frac{3}{4}\frac{\Gamfunc{n-1}{2}}{\Gamfunc{n+3}{2}}$ 
&
$\ds\frac{1}{4}\frac{\Gamfunc{n-1}{2}}{\Gamfunc{n+3}{2}}$\\[10pt]
\bottomrule
\end{tabular}
\end{center}
\caption{\label{tab:Fours} The average values of the degree four monomials 
with even exponents in $y_1,\ldots, y_{n-1}$ taken over the domain
$S(\vec{e}_n)$ with uniform probability. }
\end{table}

A similar analysis of the five monomial types appearing in the second moment's polynomial
$[\ipr(Q^{-1}\vec{y})]^2$ yields non-zero averages for those monomials of total
degree eight which we list in Table~\ref{tab:Eights}. 
\begin{table}[ht]
\begin{center}
\renewcommand{\arraystretch}{1.5}
\setlength\tabcolsep{4pt}
\begin{tabular}{@{}p{4em}ccccc}
\toprule
\textbf{Average:}
& $\expval{y_k^8}_{S(\vec{e}_n)}$ 
& $\expval{y_k^6y_\ell^2}_{S(\vec{e}_n)}$ & $\expval{y_k^4y_\ell^4}_{S(\vec{e}_n)}$ 
& $\expval{y_k^4y_\ell^2y_m^2}_{S(\vec{e}_n)}$ & 
$\expval{y_k^2y_\ell^2y_m^2y_p^2}_{S(\vec{e}_n)}$ \\
\addlinespace[0.4em]
\midrule
\addlinespace[0.5em]
\textbf{Value:}
& $\ds\frac{105}{16}\frac{\Gamfunc{n-1}{2}}{\Gamfunc{n+7}{2}}$ 
& $\ds\frac{15}{16}\frac{\Gamfunc{n-1}{2}}{\Gamfunc{n+7}{2}}$ 
& $\ds\frac{9}{16}\frac{\Gamfunc{n-1}{2}}{\Gamfunc{n+7}{2}}$ 
& $\ds\frac{3}{16}\frac{\Gamfunc{n-1}{2}}{\Gamfunc{n+7}{2}}$ 
&
$\ds\frac{1}{16}\frac{\Gamfunc{n-1}{2}}{\Gamfunc{n+7}{2}}$\\[10pt]
\bottomrule
\end{tabular}
\end{center}
\caption{\label{tab:Eights} The average values of the degree eight monomials 
with even exponents in $y_1,\ldots, y_{n-1}$ taken over the domain
$S(\vec{e}_n)$ with uniform probability. }
\end{table}

We now discuss each of the denominators appearing in these monomial integral calculations, and
do so taking the value $\Gamfunc{n-1}{2}$ into account. This quantity is
the numerator of the probability $P(\vec{x})$ of choosing points uniformly from the sphere
and is a prefactor of each moment calculation. A closed form for the first and
second IPR moments therefore depends on our ability to simplify several
$n$-dependent ratios. 

For the first moment, 
the total degree of the monomials is four, giving the denominator 
of Theorem \ref{FollandsTheorem} a value of $\Gamma(\frac{n+3}{2})$. 
Hence, we seek a closed form for 
${\Gamma(\frac{n-1}{2})}/{\Gamma(\frac{n+3}{2})}$. 
For a positive integer $k$, the Gamma function 
takes the values $\Gamma(k)=(k-1)!$ and 
$\Gamma(k+\frac{1}{2})=(k-\frac{1}{2})\cdots(\frac{1}{2})\sqrt{\pi}$ 
so that we have the following derivations:

If $n-1=2k$ is even, then 
\begin{equation}
\frac{\Gamma(\frac{n-1}{2})}{\Gamma(\frac{n+3}{2})} 
= \frac{(k-1)!}{(k+1)!} 
= \frac{1}{(\frac{n+1}{2})(\frac{n-1}{2})} 
 = \frac{4}{(n+1)(n-1)}.
\end{equation}

On the other hand, if $n-1$ is odd, then $n-1=2k+1$ for some $k$ and 
\begin{equation}
\frac{\Gamma(\frac{n-1}{2})}{\Gamma(\frac{n+3}{2})} 
 = \frac{(k-\frac{1}{2})\cdots(\frac{1}{2})\sqrt{\pi}}{(k+\frac{3}{2})(k+\frac{1}{2})
\cdots(\frac{1}{2})\sqrt{\pi}} 
 = \frac{4}{(2k+3)(2k+1)} 
 = \frac{4}{(n+1)(n-1)}.
\end{equation}

The second moment calculation contains 
only monomials of total degree eight, so that the denominator 
of Theorem \ref{FollandsTheorem} is $\Gamma(\frac{n+7}{2})$. 
In a fashion similar to the derivation above, we give 
an alternate form for the fraction 
${\Gamma(\frac{n-1}{2})}/{\Gamma(\frac{n+7}{2})}$.

When $n-1=2k$ is even, 
we have  
\begin{equation}
\ds\frac{\Gamma(\frac{n-1}{2})}{\Gamma(\frac{n+7}{2})}
= \frac{(k-1)!}{(k+3)!}
= \frac{1}{(k+3)(k+2)(k+1)(k)}
= \frac{16}{(n+5)(n+3)(n+1)(n-1)}.
\end{equation}

On the other hand, for odd $n-1=2k+1$, 
\begin{equation}
\frac{\Gamma(\frac{n-1}{2})}{\Gamma(\frac{n+7}{2})} 
= \frac{\left(k-\frac{1}{2}\right)\left(k-\frac{3}{2}\right)\cdots\left(\frac{1}{2}\right)\sqrt{\pi}}
{\left(k+\frac{7}{2}\right)\left(k+\frac{5}{2}\right)\left(k+\frac{3}{2}\right)\cdots\left(\frac{1}{2}\right)\sqrt{\pi}}
=\frac{16}{(n+5)(n+3)(n+1)(n-1)}. 
\end{equation}

\section{Evaluation of $Q$ Summations}
\label{sec:Qsums}
In this appendix we provide details on the evaluation of the summations over
the components of the rotation matrix $Q$ given in Eq.~\eqref{eq:Qcomponent}
that appear in the expressions for the first (Eq.~\eqref{eq:mu1}) and second
(Eq.~\eqref{eq:mu2sum}) moments of the of the inverse participation ratio. These
evaluations are performed by first noting that all such powers of $Q$ only
appear with the first index smaller than $n$ and in this restricted case we can
write:
\begin{equation}
    \left(Q_{ij}\right)^s = \frac{1}{(n+\sqrt{n})^{s}} \left[(-1)^s + \alpha_s(n) \delta_{nj}
    + \beta_s(n) \delta_{ij}\right]
\label{eq:Qpowers}
\end{equation}
where $\alpha_s(n)$ and $\beta_s(n)$ are power dependent functions of $n$ that
are listed in Table~\ref{tab:alphabeta} and
we have used the fact that $\delta_{ij}^s = \delta_{ij}$ and $\delta_{ij}^s
\delta_{n,j}^{s'} = 0$ $\forall s,s'\ge 1$ since $i<n$.

\begin{table}[ht]
\begin{center}
\renewcommand{\arraystretch}{1.75}
\setlength\tabcolsep{6pt}
\begin{tabular}[t]{@{}p{1.2em}p{2in}p{2.5in}}
\toprule
    $s$	& $\alpha_s(n)$ & $\beta_s(n)$ \\ 
\midrule
    1 & $-\sqrt{n}$ & $n + \sqrt{n}$ \\ \addlinespace
    2 & $n+2\sqrt{n}$ & $(n-1) (n+2\sqrt{n})$ \\ \addlinespace 
    3 & $-\sqrt{n}\left(3+3 \sqrt{n}+n\right)$ & $(n + \sqrt{n}) 
    (3 - 3 \sqrt{n} - 2 n + 2 n^{3/2} + n^2)$  \\ \addlinespace  
    4 & $(n+2\sqrt{n})(2+2\sqrt{n} + n)$ & \multirow{2}{2.6in}{$\!\begin{aligned}%
            &(n-1) (n+2\sqrt{n})\\
            &\;\times (n^2 + 2n^{3/2}-n - 2\sqrt{n}+2)
    \end{aligned}$}
    \\[1em] \addlinespace[1em] 
    6 & \multirow{2}{2in}{$\!\begin{aligned}%
            &(n+2\sqrt{n})(1+\sqrt{n}+n) \\
            &\; \times (3+3 \sqrt{n}+n) 
    \end{aligned}$}
            & 
    \multirow{2}{2.6in}{$\!\begin{aligned}%
            & (n+2\sqrt{n})(n-1)(1-\sqrt{n}+2 n^{3/2}+n^2)  \\
            &\; \times (3-3 \sqrt{n}-2 n+2 n^{3/2}+n^2)
    \end{aligned}$}
    \\ \addlinespace[2.0em] 
    8 & 
    \multirow{2}{2in}{$\!\begin{aligned}%
            & (n+2 \sqrt{n})(2+2 \sqrt{n}+n)\\
            &\; \times (2+4 \sqrt{n}+6 n+4 n^{3/2}+n^2)
    \end{aligned}$}
            & 
    \multirow{4}{2.5in}{\vspace{1em}$\!\begin{aligned}%
            & (n+2\sqrt{n})(n-1) \\
            &\; \times (2-2 \sqrt{n}-n+2 n^{3/2}+n^2) \\ 
            &\; \times (2-4 \sqrt{n}+2 n+8 n^{3/2}-5 n^2 \\
            &\qquad \; -\ 8 n^{5/2}+ 2 n^3+4 n^{7/2}+n^4)
    \end{aligned}$}\\ \addlinespace[5em]
\addlinespace 
\bottomrule
  \end{tabular}
\end{center}
\caption{\label{tab:alphabeta} Expressions for $\alpha_s(n)$ and $\beta_s(n)$
appearing as coefficients of Kronecker $\delta$-functions when evaluating
powers of the components of the rotation matrix $(Q_{ij})^s$ where $i < n$ in
Eq.~\eqref{eq:Qpowers}.}
\end{table}

\subsection{1st IPR moment}
\label{asub:1st_ipr_moment}

We begin by using Eq.~\eqref{eq:Qpowers} to perform the double and triple summations
appearing in the expression for the first moment of the inverse participation
ratio in Eq.~\eqref{eq:mu1}.
\begin{align}
    \sum_{i=1}^n\sum_{k=1}^{n-1} Q_{ki}^{4} &= 
    \frac{1}{(n+\sqrt{n})^4}
    \sum_{i=1}^n\sum_{k=1}^{n-1} \left[1+\alpha_4(n) \delta_{ni} + \beta_4(n)
    \delta_{ki}\right] \nonumber \\
    &= \frac{1}{(n+\sqrt{n})^4} \left[n(n-1) + \alpha_4(n) (n-1)  + \beta_4(n)
(n-1)\right] \nonumber \\ 
\label{eq:Qsum1}
&= 
\frac{n-1}{(n+\sqrt{n})^4}\left[ n+\alpha_4(n) + \beta_4(n)\right]
\end{align}
and using the values of $\alpha_4(n)$ and $\beta_4(n)$ in
Table~\ref{tab:alphabeta} we find
\begin{equation}
    \sum_{i=1}^n\sum_{k=1}^{n-1} Q_{ki}^{4} =  n - 
    \frac{29 + 30\sqrt{n}+5n}{(1+\sqrt{n})^2} + \frac{24}{\sqrt{n}} -
    \frac{9}{n}.
\label{eq:Q41}
\end{equation}
Now, dropping the explicit $n$ dependence of the $\alpha_s$ and $\beta_s$
functions for simplicity, the triple summation may be performed in a similar
manner:
\begin{align}
    \sum_{i=1}^{n} \sideset{}{'}\sum_{k,\ell}^{n-1} Q_{ki}^{2}Q_{\ell i}^{2} &= 
    \frac{1}{(n+\sqrt{n})^4} \sum_{i=1}^{n}\sum_{k\ne\ell}^{n-1} 
    \left(1+\alpha_2\delta_{ni} + \beta_2\delta_{ki} \right) 
    \left(1+\alpha_2\delta_{ni} + \beta_2\delta_{\ell i} \right) \nonumber \\
    &= \frac{1}{(n+\sqrt{n})^4} \sum_{k\ne\ell}^{n-1} 
    \left[(1+\alpha_2)^2 +
    \sum_{i=1}^{n-1}(1+\beta_2\delta_{ki})(1+\beta_2\delta_{\ell i})\right]
    \nonumber \\
    &= \frac{1}{(n+\sqrt{n})^4} \sum_{k\ne\ell}^{n-1} 
    \left[(1+\alpha_2)^2 + n-1 + 2\beta_2 \right] \nonumber \\
    &= \frac{(n-1)(n-2)}{(n+\sqrt{n})^4}
    \left[(1+\alpha_2)^2 + n-1 + 2\beta_2 \right] \nonumber \\
    \label{eq:Q221}
    &= \frac{(\sqrt{n}-1)(3\sqrt{n}+5)(n-2)}{n(\sqrt{n}+1)^2},
\end{align}
where we have used $\alpha_2(n)$ and $\beta_2(n)$ from
Table~\ref{tab:alphabeta}.

\subsection{2nd IPR moment}
\label{asub:2nd_ipr_moment}

There are seventeen individual summations appearing in the expression for the
average of the square of the inverse participation ratio over the sphere given
in Eq.~\eqref{eq:mu2sum} and we will include the details of only a representative
sample here.  All can be performed using similar techniques employing
Eq.~\eqref{eq:Qpowers} and Table~\ref{tab:alphabeta} and we begin with the sum
over the components of $Q^8$ which can be evaluated in exact analogy with
Eq.~\eqref{eq:Qsum1}:
\begin{align*}
    \sum_{k=1}^{n-1}\sum_{i=1}^{n}Q_{ki}^{8}  &= 
    \frac{1}{(n+\sqrt{n})^8}
    \sum_{k=1}^{n-1} \sum_{i=1}^n\left[1+\alpha_8 \delta_{ni} + \beta_8
    \delta_{ki}\right] \nonumber \\
    &= \frac{\sqrt{n}-1}{(1+\sqrt{n})^6 n^3} 
    \left(49 + 7 \sqrt{n} - 7 n + 119 n^{3/2} + 21 n^2 - 133 n^{5/2} + 9 n^3 \right.  \\
    & \left.\qquad \qquad \quad +\ 111 n^{7/2} + n^4 - 57 n^{9/2} - 13 n^5 + 13 n^{11/2} 
    + 7 n^6 + n^{13/2}\right).
\end{align*}

The next novel term includes $Q^6$ which appears as the third summation and 
is evaluated in a similar method as in Eq.~\eqref{eq:Q221} albeit with
the modification that it involves the product of two different powers of $Q$: 
\begin{align*}
 \sideset{}{'}\sum_{k,\ell}^{n-1} \sum_{i=1}^{n}  Q_{ki}^{6}Q_{\ell i}^{2} &= 
    \frac{1}{(n+\sqrt{n})^8} \sum_{k\ne\ell}^{n-1} \sum_{i=1}^{n}
    \left(1+\alpha_6\delta_{ni} + \beta_6\delta_{ki} \right) 
    \left(1+\alpha_2\delta_{ni} + \beta_2\delta_{\ell i} \right) \nonumber \\
    &= \frac{1}{(n+\sqrt{n})^8} \sum_{k\ne\ell}^{n-1} 
    \left[(1+\alpha_6)(1+\alpha_2) +
    \sum_{i=1}^{n-1}(1+\beta_6\delta_{ki})(1+\beta_2\delta_{\ell i})\right]
    \nonumber \\
    &= \frac{(n-1)(n-2)}{(n+\sqrt{n})^8} \left[(1+\alpha_6)(1+\alpha_2) +
    n-1 +\beta_6 + \beta_2 \right] \\
    &= 
\frac{\left(\sqrt{n}-1\right)(-2+n)}{\left(1+\sqrt{n}\right)^6 n^3}
\left(37+31 \sqrt{n}+10 n+40 n^{3/2}+29 n^2-15 n^{5/2} \right. \\
    & \left. \qquad \qquad \qquad  \qquad 
-14 n^3+4 n^{7/2} +5 n^4+n^{9/2}\right).
\end{align*}

The final type of term contains mixed second indices between the different
rotation matrix powers and we consider the thirteenth sum in
Eq.~\eqref{eq:mu2sum} as a representative of this set.  The strategy is the same
for all such terms and involves performing the inner summation by extracting
the terms with $i=n$ and $j=n$ and performing the summations over $j$ and $i$,
then breaking the remaining restricted sum over $i\ne j \le n-1$ into the
difference of an unrestricted sum over all values of $i,j \le n-1$ and one
with $i=j \le n-1$. We have 
\begin{align*}
    & (n+\sqrt{n})^8 \sum_{i\ne j}^{n} Q_{ki}^{3}Q_{\ell i}Q_{kj}Q_{\ell j}Q_{mj}^{2} \\ 
&\;\; = 
    \left(-1+\alpha_3\right) \left(-1+\alpha_1 \right) 
\sum_{j=1}^{n-1}
\left(-1+\beta_1\delta_{k j} \right) 
\left(-1+ \beta_1\delta_{\ell j} \right) 
\left(1+ \beta_2\delta_{m j} \right) \\
&\;\; \quad +\; \left(-1+\alpha_1\right)^2 \left(1+\alpha_2 \right) 
    \sum_{i=1}^{n-1} 
\left(-1+ \beta_3\delta_{k i} \right) 
\left(-1+ \beta_3\delta_{\ell i} \right)  \\
&\;\; \quad +\; \sum_{i=1}^{n-1}
\left(-1+ \beta_3\delta_{k i} \right) 
\left(-1+ \beta_1\delta_{\ell i} \right) 
\sum_{j=1}^{n-1} 
\left(-1+\beta_1\delta_{k j} \right) 
\left(-1+ \beta_1\delta_{\ell j} \right) 
\left(1+ \beta_2\delta_{m j} \right) \\
&\;\; \quad -\; \sum_{i=1}^{n-1}
\left(-1+ \beta_3\delta_{k i} \right) 
\left(-1+ \beta_3\delta_{\ell i} \right)^2
\left(-1+ \beta_1\delta_{k i} \right) 
\left(1+ \beta_2\delta_{m i} \right) \\
&\;\; = \left [ 
(-1+\alpha_1)(-1+\alpha_3)(n-1-2\beta_1+\beta_2)
+(1+\alpha_2)(-1+\alpha_1)^2(n-1-\beta_1-\beta_3) \right. \\
&\;\;\;\;\quad \left. +\; (n-1-\beta_1-\beta_3)(n-1-2\beta_1+\beta_2)
- (n-1-3\beta_1+\beta_2-\beta_3 + \beta_1 \beta_2 + \beta_1^2) \right]
\end{align*}
and putting everything together:
\begin{align*}
\sideset{}{'}\sum_{k,\ell,m}^{n-1} \sideset{}{'}\sum_{i,j}^n 
Q_{ki}^{3}Q_{\ell i}Q_{kj}Q_{\ell j}Q_{mj}^{2}
&= 
\frac{(1-\sqrt{n})(n-2)(n-3)}{\left(1+\sqrt{n}\right)^6 n^3}
\left(29+39 \sqrt{n}+4 n-28 n^{3/2} \right. \\
&\;\;\quad \qquad \qquad \qquad \left. -\ 12 n^2+12 n^{5/2}+10 n^3+2 n^{7/2}\right).
\end{align*}

\vspace{20pt}
\bibliographystyle{iopart-num.bst}
\bibliography{ipr_references}

\end{document}